\documentstyle[11pt,epsf]{article}

\input psfig
\def\beq{\begin{eqnarray}}
\def\eeq{\end{eqnarray}}
\def\bea{\begin{eqnarray*}}
\def\eea{\end{eqnarray*}}

\def\centeron#1#2{{\setbox0=\hbox{#1}\setbox1=\hbox{#2}\ifdim
\wd1>\wd0\kern.5\wd1\kern-.5\wd0\fi
\copy0\kern-.5\wd0\kern-.5\wd1\copy1\ifdim\wd0>\wd1
\kern.5\wd0\kern-.5\wd1\fi}}
\def\ltap{\;\centeron{\raise.35ex\hbox{$<$}}{\lower.65ex\hbox{$\sim$}}\;}
\def\gtap{\;\centeron{\raise.35ex\hbox{$>$}}{\lower.65ex\hbox{$\sim$}}\;}

\def\singleandthirdspaced{\baselineskip=\normalbaselineskip\multiply
    \baselineskip by 130\divide\baselineskip by 100}
\def\singlespaced{\baselineskip=\normalbaselineskip}

\newcommand{\newc}{\newcommand}
\newc{\qbar}{{\overline q}}
\newc{\Kahler}{K\"ahler }
\newc{\deltaGS}{\delta_{\rm GS}}
\begin{document}
\begin{titlepage}
\begin{flushright}
{\large
hep-th/9903019 \\
SCIPP-99/03; RU-98-53; UW/PT-99-05\\

}
\end{flushright}

\vskip 1.2cm

\begin{center}

{\LARGE\bf Constraints on Theories With Large Extra Dimensions}

\vskip 1.4cm

{\large
Tom Banks$^a$, Michael Dine$^b$, and Ann E. Nelson$^c$}
\\
\vskip 0.4cm {\it $^a$Physics Department, Rutgers University,
Piscataway, NJ  08855-0849}
\\ {\it $^b$Santa Cruz Institute for Particle Physics,
     Santa Cruz CA 95064  } \\
{\it $^c$Physics Department,
     University of Washington,
     Seattle, WA  98195-1550} \\

\vskip 4pt

\vskip 1.5cm

\begin{abstract}
Recently, a number of authors have challenged the conventional
assumption that the string scale, Planck mass, and unification
scale are roughly comparable.  It has been suggested that the
string scale could be as low as a ${\rm TeV}$.
In this note, we explore constraints on these
scenarios.  We argue that the most plausible cases have a
fundamental scale of at least $10$ TeV and five dimensions
of inverse  size $10 $ MeV.  We show that a radial dilaton mass in the
range of proposed millimeter scale gravitational
experiments arises naturally in
these scenarios.  Most other scenarios require huge values of
flux and may not be realizable in M~Theory.
Existing precision experiments put a
conservative lower bound of $6-10$ TeV on the fundamental energy
scale.  We
note that large dimensions with bulk supersymmetry might be a
natural framework for quintessence, and make some other tentative
remarks about cosmology.
\end{abstract}

\end{center}

\vskip 1.0 cm

\end{titlepage}
\setcounter{footnote}{0}
\setcounter{page}{2}
\setcounter{section}{0}
\setcounter{subsection}{0}
\setcounter{subsubsection}{0}

\singleandthirdspaced

\section{Introduction}

It has been traditional, in string theory, to assume that the
compactification scale and the string scale are both large,
comparable to the Planck scale.  This stemmed from the belief that
sensible string phenomenology could only emerge from the heterotic
string.  In weakly coupled heterotic string theory, the
coincidence of these scales is inevitable, since the observed four
dimensional gauge couplings are inversely proportional to the
volume of the compact space, $V_d$.  With the recent discovery
that all string theories are different limits of a larger theory,
called M~Theory, these assumptions have been called into question.
Indeed, they do not even hold in weakly coupled Type I theory.
Witten argued that the phenomenon of coupling constant unification
suggests that the proper description of nature might be as a
strongly coupled heterotic string, in which the fundamental Planck
scale was of order the unification scale, with the eleventh
dimension  about $40$ times larger \cite{wittencompact}.
More generally, it has been realized that in M~Theory, nonabelian gauge
fields arise on lower dimensional submanifolds of the internal space,
which have been named {\it branes}.  It is thus quite natural
to have an internal manifold somewhat larger than the fundamental scale,
which would then give a large effective four dimensional Planck mass,
without affecting the gauge couplings.

Ref. \cite{bdscales} pointed out a difficulty with this sort of
picture.  In a theory where at least five dimensions are large, it
is difficult to understand the stabilization of the radii.  If the
underlying theory is supersymmetric, the bulk supersymmetry, which
is at least that of a five dimensional theory, requires that the
potential vanish for large radii.  This means that the radii must
be stabilized at values of order the fundamental scale.  This is
similar to the argument that the gauge coupling must be of order
one.   In the case of the gauge coupling, the problem is to explain
why the coupling is $1/25$ instead of one, and why a weak coupling
expansion should hold in the low energy theory.  For the case of
the scales of a supersymmetric theory, one must understand why
there are ratios of scales of order $40$ rather than $1$, and why
a large radius $M$-theory type picture is valid.  A possible
explanation why numbers normally of order one might be $10-100$
was described in \cite{coping} and \cite{bdscales}.  
Note also, that in Witten's picture there is only one large dimension,
whereas in general there might be as many as seven.  This would further
reduce the size of the internal manifold necessary to the explanation of
the discrepancy between the unification and Planck scales.

While explaining numbers of order $10-100$ may be troubling, the
only enormous hierarchies in these pictures which require
explanation are the ratio of $M_w$ over $M_p$, and the smallness
of the cosmological constant. The first problem is traditionally
explained
through exponentially small low energy field theory effects, and
the latter problem is a feature of any description of the world in
terms of low energy effective field theory \footnote{As it is
conventionally understood.  See however \cite{ckn,horava}.}.

By contrast, the radii, in units of the fundamental Planck scale
or string tension, are at most numbers of order a few (perhaps
$70$ in the Witten proposal, and smaller in generalizations of it
with more dimensions a bit bigger than the GUT scale), and the
basic couplings are numbers of order one. The coincidence of the
radial size with the GUT scale allows us to understand the
evidence for coupling unification and neutrino masses as probes of
the fundamental scale of the theory, while the Planck scale arises
from the product of a large number of numbers of order one, and
has no fundamental significance\footnote{We note also that the
simplest way to understand the magnitude of density fluctuations
as measured by COBE, invokes an inflationary potential of the form
$M_{GUT}^4 v(\phi / M_P).$  This suggests again that $M_{GUT}$ is
the fundamental scale and that the inflaton is a bulk modulus in a
brane scenario with SUSY broken only on branes.}. The high
fundamental scale also makes it easier to suppress violations of
baryon number, lepton number and flavor, though in a theory with
standard model superpartners around the TeV scale, one also needs
discrete symmetries to ensure that these approximate conservation
laws are preserved with the requisite accuracy.

More recently, a number of authors have considered the possibility
that the compactification (energy) scale is far lower, with the
fundamental scale of string theory being as low as a
TeV\cite{lykken,dimopoulos,braneworld,precursors,otherextradim}.
They argue, first, that
this might be desirable. Scalar masses would then naturally be of
order $1~ {\rm TeV}$ or so, eliminating the conventional fine
tuning problem associated with the Higgs boson. Proton decay could
be forbidden by discrete symmetries, much as in the more
conventional supersymmetric case, though more intricate symmetries
may be required (bulk gauge symmetries were suggested in
\cite{savasflavor}). Flavor problems can be suppressed if one
supposes that there are sufficiently elaborate non-abelian flavor
symmetries. More serious issues are posed by the production of the
light Kaluza-Klein (KK) states.  But these authors show that
provided the compactification radius is smaller than a millimeter,
then production of these particles is not a serious
problem\cite{savasconstraints}. Accelerator experiments set limits
in the ${\rm TeV}$ range\cite{accelerators}, while astrophysics in
some cases sets stronger limits\cite{savasconstraints}.  In
particular, in the scenario with two large dimensions, where KK
states might be observable in sensitive gravitational experiments,
the bounds on the higher dimensional Planck scale are pushed up to
$30$ TeV.

The possibility that we might observe violations of
Newton's law of gravity in  sensitive tabletop
experiments, or probe extra dimensions in accelerators
is extremely exciting.  The fact that such a possibility
is not a priori ridiculous is very interesting.
But it is natural to ask:  how plausible is it that the
extra dimensions be so large?

In this note, we will address this question from several
points of view.  We should stress from the start that
we will not be able to demonstrate that a low string
scale and large compact dimensions
are impossible.  We will concentrate on scenarios in
which the fundamental scale is within the reach of planned accelerator
experiments.  Some of our remarks and techniques may be applicable
to the analysis of other parts of the wide spectrum of choices which
have been explored for the sizes of extra dimensions \cite{otherscen}
but we will not discuss these more general scenarios.  Our discussion of
laboratory constraints is definitely not relevant to most of these more
general scenarios.

Our discussion will focus on three issues:
\begin{itemize}
\item  The problem of fixing the moduli:
This problem takes on different forms according to the manner in
which SUSY is broken.  There are two ways to break SUSY consistent
with observation.  It can be broken (at or near the
fundamental TeV scale) on the brane where the standard model
lives, but unbroken in the bulk.  Or it can be broken everywhere
at the fundamental scale. In the second case, there is an
essentially unique way to stabilize the radial dilaton, which is
to balance the energy of a large flux against the bulk
cosmological constant, as proposed by Arkani-Hamed,
Dimopoulos and March-Russell\cite{savasradii}, following earlier
suggestions by Sundrum\cite{sundrum}.  (Other discussions of
stabilization appear in \cite{morestabilization}.)
We show that in this case, the requisite fluxes are huge, always
larger than $10^5$.  When SUSY is
(approximately) preserved in the bulk,
the mechanisms of stabilization are more intricate.
In many cases the radial dilaton mass is extremely
small, and it is likely to lead to observational problems.
However, there is a scenario with $5$ relatively large dimensions
and a fundamental scale of order $10$ TeV, where the flux needed
to stabilize the radius is of order one.\footnote{We must also invoke a
plausible weak coupling factor and a scale of SUSY breaking of order $1$
TeV on the brane.}  In this case the radial dilaton is within range of
proposed improvements of the Cavendish experiment.

 We
emphasize however that in most of these models,
the hierarchy problem is solved in
the low energy effective theory by introducing a large integer
valued conserved flux on a compact manifold. There is no guarantee
that M Theory allows such large fluxes, and there are no known
supersymmetric vacua with flat noncompact dimensions which support such
large values of the flux.

Finally, we note that if we do not stabilize the
radial dilaton in the SUSY case, we might be led to an
interesting model of quintessence\cite{quintessence,carroll,koldalyth,
riotto,benakli} This
possibility is potentially exciting for several reasons.  First,
it might provide a natural understanding of the large value of the
radius.  In addition, the scale of supersymmetry breaking and of
the vacuum energy are not so closely tied as in more conventional
pictures. As we will explain, quintessence based on bulk moduli
fields in a brane world does not have a problem with the time
variation of the fine structure constant.   Unfortunately, bounds
on the time variation of Newton's constant require that the radial
dilaton did not change drastically during post BBN cosmological
history (so that a cosmological explanation for the large size of the
extra dimensions must involve cosmology before this time).
In addition, the radial dilaton is extremely light, and
generically possesses Brans-Dicke couplings of order one, which will
cause problems for astronomy.

\item  Problems of flavor:  Here we include not just flavor, but
also the effects of higher dimension flavor-conserving operators.
These suggest lower limits of order $6\ {\rm TeV}$ on the scale of
the theory. As explained in ref. \cite{savasflavor,dvaliflavor},
one can hypothesize mechanisms for fermion mass generation which
lead to suitable textures, and these mechanisms can suppress many
dangerous flavor-changing processes.  There are  two difficulties,
however, one theoretical and one phenomenological.  The flavor
proposals of \cite{savasflavor,dvaliflavor} require that flavor
symmetries be broken on ``other, faraway'' branes from the one on
which the standard model fields live.  But, as we will explain,
given the lack of supersymmetry one expects that there is a
potential between the branes, and they are not likely to be widely
separated. This flaw, as we will see, is hardly fatal, but it is
still one more coincidence required for the whole picture to make
sense. Second, these mechanisms do not account by themselves for
the smallness of $CP$ violation. Without supposing additional
structure or very weak couplings, one obtains a limit of at least
$10~ {\rm TeV}$. These scales are already troubling from the point
of view of scalar masses.  One needs to explain why the Higgs
mass-squared is $100$ to $1000$ times smaller than its expected value. This
criticism is certainly not decisive, and it also applies to many
supersymmetric models.
\item  Cosmology.  Here our statements will be quite tentative.
We know little about the cosmology of such high dimension
theories. Still, there are several puzzles.  Most of these are
connected with the question of how the theory finds its way into
the correct vacuum.   Essentially all of them lead back to an
inflationary era\cite{morecosmo,lyth,lindekaloper,ovrut}.  Inflationary cosmology
in a brane world scenario has not been sufficiently explored to warrant
definite conclusions.  We will simply exhibit the
formidable problems to be solved and make a few remarks
about attempts to solve them.   In particular, in arguing that scenarios
with millimeter extra dimensions were compatible with Big Bang
Nucleosynthesis, the authors of \cite{savasconstraints}  assume initial
conditions in which the brane is excited to about 1 MeV, while bulk
modes of much lower energy are in their ground state.  We suggest that
it may be very difficult to find plausible conditions at higher energy
which leave the system in this rather bizarre state.  In particular,
although Dvali and Tye \cite{dvalitye} have proposed an explicit model
in which a field on the brane dominates the energy density at some point
in time, the reheat temperature they obtain is too high, and the bulk
will become overexcited in this model.  We suspect that this is a quite
general feature.  In addition we point out that models with only two
large dimensions have much stronger couplings between brane and bulk,
since homogeneous excitations on the brane give rise to fields which do
not fall off in the bulk.
Our conclusion is that models with a KK threshold somewhat above an
MeV are much more likely to be viable.

\end{itemize}

\section{Fixing the Moduli}

It is natural to divide this discussion into two parts, according
to the manner in which SUSY is broken.  Note that supersymmetry is
almost certainly broken on the brane at a scale comparable to a
$\rm TeV$.  A much lower scale for SUSY breaking would bring us into
conflict with direct  searches.
This however still leaves us
the option of preserving SUSY in the bulk.  The stabilization
problem has a rather different character in this case.  We begin
with the case where SUSY is completely broken in the bulk as well
as on the brane.

Our treatment of the non-supersymmetric case differs from that of
 \cite{savasradii} in the way in which we treat curvature and
the cancellation of the effective cosmological constant.
The latter is the cosmological constant
in the effective theory below the KK energy scale and is the parameter
 actually measured by observation.  It is the sum of a bulk term, a
 boundary term and a radiative correction term.
In \cite{savasradii} it was assumed that the bulk cosmological constant
 was tuned to cancel the boundary cosmological constant.  For large
 radius this implies that the dimensionless coefficient of the bulk term
 is extremely small, of order $(RM)^{-n}$, where $n$ is the
number of large compact dimensions.
The boundary cosmological constant is of order
 $M^4$, with $M \sim 1$ TeV in the most ambitious models.
In this case it is consistent to neglect the
 curvature terms in the effective action because the curvatures obtained
 by solving the Einstein equations in the presence of the bulk
 cosmological constant are smaller than or
 equal to the cosmological term itself.  The resulting bulk geometry is
 approximately flat and the KK modes have masses of order $1/R$.  The
 Compton wavelength of the radial dilaton is of order a mm. independent
 of the number of spatial dimensions.  This follows from the fact that
 the dilaton is a bulk modulus and the overall scale of its potential is
 set by the boundary cosmological term.\footnote{This conclusion
follows from equation (2.26) of \cite{savasradii} but was not
emphasized in that paper.}

We believe that a more plausible mechanism for fine tuning the effective
cosmological constant is obtained by allowing the curvature terms in the
effective action to have the order of magnitude suggested by dimensional
analysis.  Then one can fine tune the cosmological constant by
canceling the bulk cosmological term against the leading curvature
term.  The coefficient of the bulk term now only needs to be a factor
of $(RM)^{-2}$ smaller than its naive expectation, in any number of
dimensions.   Of course, since we are talking about a single fine tuning
in either case, the reader may feel that our choice is purely a matter
of taste.  We do not have strong arguments against this position, but
will nonetheless present our results with the curvature terms in place.

We argue
that the only plausible mechanism for stabilizing
a large radius is to balance a large flux
against the bulk cosmological constant, and curvature terms, as suggested
in \cite{savasradii}.  In this case the internal manifold
will not be Ricci flat.   Thus we find that the
only case with Cavendish signals is that with
two dimensions of millimeter scale.  This scenario
requires a huge flux, which although technically
natural, does not seem to be a likely vacuum state
for M theory.   We also find that the natural scale for the bulk
vacuum energy, and thus for the flux which stabilizes the radius,
is much larger than that found in \cite{savasradii} in all dimensions
above two.

 Our treatment of the supersymmetric
case is quite different from that of \cite{savasradii}.
In particular, it is not
correct that SUSY breaking in the brane induces a small bulk
cosmological constant. The potential always falls to zero at
infinite $R$ in the SUSY case.  We find an acceptable value for the
dilaton mass only for rather large $n$.  By raising the fundamental
scale, and invoking SUSY broken at the TeV scale on the brane, we find a
scenario with only modest values of the fluxes, and a radial dilaton
within reach of gravitational experiments.  This model has five
dimensions, whose inverse size is of order $10$ MeV.

\subsection{Non-Supersymmetric Bulk}

We will concentrate on a particular modulus, called the radial
dilaton, $R$, that parameterizes the overall scale of the internal
geometry.
In order to study the regime of very large radial dilaton $R$
we can use the techniques of low energy effective field theory.
The leading term in the large $R$ expansion of the effective potential
for $R$ is the bulk cosmological constant: $\Lambda^{n+4} R^n$, where
$n$ is the number of large compact dimensions.
{}From dimensional analysis, we expect that $\Lambda^{n+4} =a M^{n+4}$,
with $a$ a constant of order unity \footnote{The appearance of a bulk cosmological constant raises the specter
of inflation in the extra dimensions.  The authors of reference
\cite{savasradii} dealt with this by insisting that the
Hubble radius corresponding to the cosmological constant was
always larger than the actual radius of the internal dimensions.
This constraint can be understood in another way. If we consider
solutions of the field equations involving time dependent moduli
of the internal dimensions, then the moduli appear as scalar
fields in a lower dimensional theory.  In other words, we can
imagine integrating out the massive Kaluza-Klein modes and
obtaining an effective action for the moduli. If the potential for
these fields (including the cosmological constant term) has a
stable minimum then we have a static solution, with no inflation
or other evolution. If we wish to treat $\Lambda_b$ as a small
parameter, consistency requires that the shifts in the massive
modes, as we turn on $\Lambda_b$, be small. It is a simple
exercise to show that if $\Lambda_b < M^{n-2} /R^2$, the
contribution to the mass of the Kaluza-Klein states from the
cosmological term is small, and the shifts are small ($\Lambda_b
R^4 < R^2$). This question leads to precisely the constraint found
by Dimopoulos et al., $\Lambda_b < M^{n+2}/R_o^2$. We do not see,
however, that this constraint is truly necessary.
Once $\Lambda_b
> 1/R^2$, the cosmological term makes a large contribution to the mass
of the low-lying Kaluza-Klein states, but there
are perfectly good static solutions of the equations
of motion.  The naive idea that internal dimensions
of scale $R$ would lead to particles of mass $~ 1/R$
is incorrect in this case, and the phenomenological
consequences would be different.  However, we will
see below that the requirement that the four
dimensional effective cosmological constant comes
out consistent with observation, requires the
bulk cosmological constant to be small enough that
this issue never arises.}.

Higher order terms in the expansion of the effective potential 
for large $R$ can come from three
sources.  As noted in
\cite{savasradii}, if the internal space has non-zero
Ricci curvature, one gets a term behaving as
$M^{n+2}R^{n-2}$. Higher
orders in the curvature and its covariant
derivatives can also be important if $n$ is large
enough.  The next to leading term is of order
$M^n R^{n - 4}$, and comes from terms quadratic in curvature.  Even if the
curvature vanishes, as on a torus, there is a
Casimir energy of
the massless modes. It is of the form $c / R^4$, with a
coefficient $c$ which is of order one and may be positive or
negative.  Other terms,  which scale as various
powers of $R$, can
occur if the low energy effective field theory contains
antisymmetric tensor gauge fields. From a flux $Q$ of a field with
rank $p$ field strength tensor, threading a $p$-cycle of the
internal manifold, we get an energy $ \sim e^{-2} Q^2 M^4 (MR)^{n -
2p}$.  Here $e$ is a dimensionless coupling which in principle
could be large or small.  In M~Theory it would be a modulus.
Thus, a truly large or small value of
this parameter would require an explanation.
We would again argue that the potential for this
modulus would be unlikely to have a minimum at such
extreme values.

If all dimensionless coefficients are of order one, and
there is a stable minimum for the potential
described above,
then it occurs for $R \sim M^{-1}$ and gives a four dimensional
cosmological constant of order $M^4$.  However, we know that
effective field theory generally gives a large value of the
cosmological constant, so the simpleminded dimensional analysis
argument may well be incorrect.

A conservative, or at least conventional, way to deal with this
problem is to assume that the effective four dimensional
cosmological constant vanishes (or takes on a tiny nonzero value
consistent with observation) by some mechanism unknown in
effective field theories, and pursue the consequences of this
constraint on the rest of the physical problems at hand.

This viewpoint places a strong upper bound on the bulk
cosmological constant. The effective four dimensional cosmological
constant is 
\beq \Lambda_4 = \Lambda_{4+n}R^n + d M^{n+2} R^{n-2}
+ \Lambda_{boundary} + {\rm radiative}\ {\rm corrections} \ .
\eeq
Here,  $d$ is a numerical constant, and the boundary cosmological
constant is a term $\Lambda_{boundary} \sqrt{-g_{ind}}$, where
$g^{\mu\nu}_{ind}$ is the metric induced on the wall by the metric
in bulk. Naive dimensional analysis estimates it to be of order
$M^4$, and we will assume that it is of this order in most of what
follows. The authors of \cite{savasradii} have proposed a ``brane
crystallization'' mechanism for stabilization in which it might be
much larger.  The idea is to have a large number of branes, each
with tension of order $M^4$.  We do not fully understand this
scenario (in particular, we do not understand how Gauss' law can
be satisfied in the presence of a vast number of branes with the
same charge)
 and will not explore it here. SUSY on
the brane must be broken near the TeV scale, in a
phenomenologically viable theory, so a small boundary cosmological
constant would definitely be an extra fine tuning.

With the conventional sign for the Einstein action, a manifold
with positive integrated Ricci curvature, will make a negative
contribution to the energy. If the bulk cosmological constant is
positive, we can obtain a cancellation of the effective
cosmological constant between these two terms. In order for this
to occur, the dimensionless coefficient of the bulk cosmological
constant must be very small if $R_0 M$ is very large ($R_0$ is the
value of the radial dilaton at the minimum of the potential),
$\Lambda_{4+n} \sim (R_o M)^{-2}M^{4+n}$. Of course, the full
cancellation of the effective cosmological constant requires us to
take into account many terms in the effective potential. This will
involve further fine tuning of the coefficient of the bulk
cosmological constant but will not change the order of magnitude
estimate of its size.

We can obtain a similar cancellation with a
negative integrated Ricci curvature and a
negative cosmological constant.  Note however
that if both the integrated curvature and the bulk
cosmological term contribute to the potential
with the same sign,and if $n>2$,
 then we can obtain a
cancellation at large $R_0 M$
only when both of these terms are much smaller
than indicated by dimensional analysis.
We consider such a double fine tuning unacceptable
and will not consider such scenarios further.

The careful reader may wonder why we have not
considered a cancellation of the cosmological
constant which involves a bulk term of natural
order of magnitude, and some term lower order in
$R$ with a very large coefficient (we will encounter
such a term in a moment).  Such a cancellation
is possible, but the resulting mass for the
radial dilaton and KK modes is very large
and most of the phenomenology expected for large
extra dimensions disappears in this scenario.

One might have imagined that, having made
the two largest terms in the effective action
of the same order of magnitude, that we could
obtain a suitable minimum just from these two terms.
This is not the case.  It is easily verified
that the relevant variational equations are
the Euclidean Einstein equations with
a cosmological constant and that the action
(which is the energy in the non-compact dimensions)
is a negative number of order the cosmological
constant.  Thus, the cancellation of the effective
cosmological constant is incompatible with solving
the equations with just these two terms.

Thus, some other term must come into the solution
of the variational equations, which means that this
term must have a coefficient much larger than
is expected on the basis of naive dimensional
analysis.  There is a unique, technically natural
way to obtain such a term. One can have a large
p form flux, $Q$, wrapped around some p cycle of the
internal manifold.  The corresponding effective
potential is \beq V(R) = M^4 [a (MR_0 )^{-2}
(MR)^n  - b (MR)^{n-2} +
{e^{-2} Q^2 \over (MR)^{2p -n} }], \eeq where $a$ and
$b$ are positive constants of order one.

The equations for a minimum with zero cosmological constant (up to
corrections sub-leading in $MR_0$) may be written as : \beq e^{-2}
Q^2 = (MR_0)^{2p - 2} (b - a) , \eeq and \beq (2p - 2) (b - a) =
2a .\eeq Thus we must have $b > a > 0$ and $p > 1$ in order to
satisfy these equations. Since $p$ is an integer, this means that
$e^{-2} Q^2$ is always greater than $(M R_0)^2$. Remembering that
$M R_0 \sim (M_P / M)^{2/n}$ we find that even for $M \sim 10 $
TeV\footnote{We choose this scale both because it is of order the
phenomenological bounds we derive in the next section, and because
it was claimed in \cite{savasradii} that one could obtain a
minimum with $6$ large dimensions and flux of order one when the
fundamental scale is 10 TeV.  Their claims were based on the
assumption of a very small bulk cosmological constant.} and seven
large dimensions,the flux is of order $10^4 $.  Note further that
eleven dimensional SUGRA does not have a two form field strength.
Thus, in M~Theory, we have $p = 4$ and the flux would be of order
$10^{13}$ in seven large dimensions. We can do somewhat better in
string theoretic limits in six large dimensions.  Heterotic and
Type I and IIA strings have bulk two form field strengths and
yield suitable $R_0$ with fluxes of order $10^5$.  In this case
one might hope to do better if the dimensionless coupling $e^2$ is
small. In Type IIA theory the two form is a Ramond-Ramond field
and we do not expect such a factor, but in the heterotic/Type I
theories we would get an enhancement for magnetic fluxes at weak
coupling. Note however that if we make this parameter very small
we are introducing another stabilization problem.  The string
coupling is a modulus and it is difficult to understand how it is
stabilized at a very small value.  We believe that it is
unreasonable to expect $e^2$ to be smaller than $10^{-2}$.  To
obtain values even this small from string dynamics , one has to
invoke the ill understood notion of Kahler stabilization
\cite{coping}. We also note that in these scenarios, independent
of the dimension, the radial dilaton mass is of order $1/ R_0$.

A word should be said about the other moduli of the internal space,
which we have been presuming fixed.  In general, if $p < n$ and the flux
is very large, this assumption is probably untenable.  The potential for
the other moduli will be overwhelmed by a huge flux which really wants to
blow up only one $p$ cycle of the manifold.  On a torus one can \lq\lq
solve \rq\rq this problem by putting flux on a complete set of cycles
({\it e.g.} put two form flux on both the $12$ and $34$ cycles of a four
torus).   It is not clear that this is possible on a general curved
manifold.   Thus, it may be that the case $p=n$ is the only one where
the analysis we have made above really works.  Other values of $p$ might
lead to fine tuning problems for the potentials of moduli other than the
radial dilaton.  We will continue to ignore this problem, but we
consider it a further indication of the delicate balancing act which
must be performed to find a theory with large dimensions.

Let us also discuss briefly the scenario of \cite{savasradii} in which
the dimensionless coefficient in the bulk cosmological term is taken so
small that the term is of the same order as the boundary cosmological
constant.  In this case, the overall scale of the potential is $M^4$,
and for $M$ of order a TeV, the radial dilaton Compton wavelength is
within reach of proposed improvements of the Cavendish experiment, for
any number of large compact dimensions.  The cosmological constant is so
small that it does not affect the masses of the KK modes.

Actually, one should be somewhat careful in the two dimensional case.
The brane tension (boundary cosmological constant) is enhanced by a
factor of $ln\ (MR_0 )$ because long range fields do not
fall off at infinity.  This, via the fine tuning of the effective
cosmological constant, in turn enhances the bulk cosmological term.
As a consequence, the masses of both radial dilaton and KK modes are
enhanced by a factor of $\sim \sqrt{ln\ (MR_0 )} \sim 7$ .  Thus, one
cannot probe the extra dimensions until one gets up to these higher
energy scales.  Combined with the lower bounds on the fundamental
scale which we will derive in the next section, this result may be
discouraging for the prospects of testing the two dimensional scenario
in experiments on gravity at millimeter scales.

As we have emphasized, for $n > 2$ it seems likely that the curvature
terms will be important.  In their presence, the cancellation of the
effective cosmological constant implies a much larger bulk energy
density than in the almost flat case we have just discussed.
In these curved scenarios we find a radial dilaton mass of the same
order as the KK mass, which  is of order its naive flat space value
$1/R_0$.  Thus the estimates of production cross sections and discovery
criteria cited in \cite{accelerators} will not be substantially changed
by our reevaluation of the stability of large dimension scenarios.
 One concludes then that effective field theory analysis shows that the
only way to stabilize the system at large $R_0 M$ is to have a large
integer flux, as originally proposed by \cite{savasradii}.
Our analysis differs from theirs (for $n>2$) by
including curvature terms for the internal manifold,
which dominate the effective potential.  As
a consequence, even a large number of large
dimensions, with a $10$ TeV fundamental scale
still require integer fluxes of order $10^5$.
The hierarchy problem is to a large extent
solved by hand.

In the low energy effective theory, large fluxes
appear to be technically natural because they
are quantized and conserved.  At a more microscopic
level one should
investigate the possibility that
there are quantum processes which can
screen the flux by popping branes out of the vacuum.
For M~Theorists, the plausibility of this scenario
becomes a question of
whether vacuum states with large flux and flat
external dimensions exist.  We know of no
SUSY vacuum
states with this property.  Fluxes have a tendency
to appear in Chern Simons like terms and to be
bounded by considerations of anomaly cancellation.
However, the present subsection is devoted to
non-supersymmetric vacuum states and our
understanding of those is practically nil.

\subsection{Supersymmetric Stabilization and
Quintessence}

If we assume that the system becomes supersymmetric in the large
radius limit, then the story is quite different.  In this case the
bulk cosmological constant vanishes.  Many of the terms in the potential
which determine the value of the radius, are now {\it a priori}
much smaller than the constant which determines the four
dimensional cosmological constant.

It is also natural to impose a Ricci flatness condition on the bulk
geometry, since this is the simplest way to preserve SUSY in the low
energy theory.  To be specific, we might imagine a IIB string
or F theory
compactification which preserves eight supercharges except on a threebrane
(which is point-like in the compact dimensions) where various gauge
fields live.  The remaining four supercharges on the brane might be
broken dynamically by gauge dynamics.  The only unnatural thing about
this idea in the context of a very low fundamental scale is the fact
that we want the SUSY breaking scale to be close to the fundamental
scale, so the treatment of the SUSY breaking dynamics by low energy
effective field theory may be suspect.  Since we are using this scenario
merely to fix ideas, we will not explore this issue.

The leading terms in the effective action that cannot be set to
zero by making technically natural discrete choices, now come from
curvature squared terms.  They are of order $M^4 (MR)^{(n-4)}$.
The overall coefficient of this term might be either positive or
negative, but is naturally of order one.  As before, the only way
to achieve a technically natural stabilization at large radius is
to introduce a term with large flux which is sub-leading in the
large $MR$ expansion.  This means that $p > 2$. We note that in
the presence of flux, the manifold is no longer Ricci flat, but
the additional Ricci tensor term scales just like the flux term
which produced it.  Thus it does not change the argument
substantially.

The details of the stabilization depend crucially on whether $n < 4$.
If it is, then the fine tuning of the $4$d effective cosmological constant
does not require any parameter in the higher dimensional Lagrangian
to take on a particularly small value.   Stabilization is decoupled
from the problem of the cosmological constant.  To achieve it, the
sign of the $(MR)^{(n-4)}$ term coming from the various quadratic
curvature invariants must be negative.  Furthermore, $e^2 Q^2 \sim (R_0
M)^{(2p -4)} \sim (M_P/M)^{2 (2p -4)/n}$.   The smallest flux is
obtained for $p=3$ and $n=3$ , and is $10^{10}$ when $M \sim 10$ TeV.

The mass of the radial dilaton is much smaller in these supersymmetric
scenarios.  For $n < 3$ the formula is \beq m \sim M (M/M_P)^{(4/n)}.
\eeq  Even for a scale of $10$ TeV and $n=3$, this is of order
$10^{-8}$ eV and the force coming from exchange of this particle should
have been seen in existing experiments.  Note that SUSY will imply that
to leading order in the $1/(R_0 M) $ expansion, the couplings of this
particle are universal.

When $n=4$ the terms quadratic in curvature are $R$ independent
and are not useful in the stabilization program.  The fine tuning of the
observed cosmological constant again has no effect on the relevant terms
in the potential\footnote{This is because it can be viewed as the fine
tuning of a much larger, but $R$ independent, term in the potential.}, 
which are now cubic curvature terms (scaling like
$(MR)^{-2}$, and a large flux term with $p > 3$ (there are also terms
with covariant derivatives of curvature which have the same scaling as
the cubic ones).   The conditions for stabilization give
$e^2 Q^2 \sim (R_0 M)^{2p - 6} \sim (M_P / M)^{(p-3)}  $.  The smallest
value is obtained for $p=4$ and equals $10^{15}$.
The radial dilaton mass comes out of order $M (M/M_P)^{3/2}$.  Even for
$M$ of order $10$ TeV this gives a Compton wavelength of order
a kilometer.  Thus, this scenario is ruled out by existing gravitational
experiments, since there is no way to tune the radial dilaton couplings
to be much smaller than that of gravity.

For $6 \geq n >4$, the quadratic curvature terms in the effective action
give rise to terms in the potential which grow with $R$.  If we allow
this term to be of its natural order of magnitude, then the whole extra
dimension picture is modified.  Dimensions of size $R$ no longer give
rise to KK states with masses of order $1/R$ because the large term in
the action gives a large mass to all of the modes.  Furthermore, the
next largest term in the inverse $RM$ expansion is the $R$ independent
boundary cosmological term.  Thus, the coefficient $c$ in the effective
potential term $c M^4 (RM)^{(n-4)} $ should be of order $(R_0
M)^{(4-n)}$ (where $R_0$ is the value of $R$ at which the potential is
minimized ) in order to obtain the right order of magnitude for the
observed cosmological constant.  This is of course a fine tuning, but
since it is the same fine tuning which sets the cosmological constant,
we should discount it.  A technically natural minimum at large $R$
can be obtained by adding a large flux term , $Q^2 M^4 (RM)^{(n - 2p)}$
with $p > 2$ ($p > 3$ if $n=6$),
to the action.  Stabilization can be achieved if
$Q \sim (R_0 M)^{(2p -n)/2} \sim (M_P /M)^{(2p/n - 1)}$.  For $p=3$ and
$n=5$ this is of order $10^3$, the smallest value we have yet seen, when
the fundamental scale is of order $10$ TeV.
Indeed, if the coupling of the three form gauge field strength is as
weak as a standard model coupling, the amelioration of the flux bound
mentioned above is likely to make the necessary integer of order $100$,
which is perhaps more palatable than the other examples we have found.
If the three form is a Neveu-Schwarz field strength in weakly coupled
IIB string theory, then a large weak coupling factor is natural.
One would have to explain the stabilization of the string
dilaton in this regime, however.

For this range of $n$ the magnitude of the effective potential is of
order $M^4$ for all $n$, so the radial dilaton mass is of order $M^2 /
M_P$ for all scenarios in this group.  Thus, if $M$ is of order $1$ TeV
we have a Compton wavelength in the range of proposed improvements of
short distance tests of gravity.  In particular, this is true for
the attractive scenario with $p=3$ and $n=5$.

We can make things even better by invoking SUSY on the brane.
Suppose the fundamental scale $M$ is $10$ TeV, while SUSY is broken
on the brane at around $M_{SUSY} \sim 1$ TeV.  Then the equation for
$Q$ is replaced by $ Q \sim (M_P / M)^{(2p/n - 1)} (M_{SUSY} / M)^2$,
because the scale of the potential will be lowered to $M_{SUSY}^4$.
If we take $p=3$ and $n=5$ as above, we get $Q \sim 10$.
Taking into account the weak coupling factor mentioned above, this could
be an integer flux of order one.   Note further that the radial dilaton
mass in this case would be in the range of Cavendish experiments.  As
far as we can see, this is the model with the fewest large dimensionless
numbers in it, which still gives exciting near term phenomenology.

For $n=7$ both the quadratic and cubic curvature terms in the effective
action give terms in the potential which
grow with $R$.  In order to cancel the cosmological constant with
only one fine tuned parameter, we must take the dimensionless
coefficient of the quadratic term to be of order $(R_0 M)^{-2}$.
In order to find a stable minimum, we must add a large flux term of the
form $Q^2 M^4 (RM)^{7 - 2p}$ with $ p \geq 4$. The only value of $p$
expected from eleven dimensional SUGRA is $ p = 4$.  This also gives the
lowest value of flux, namely $Q^2 \sim (R_0 M)^{(2p - 6)}$ or $Q \sim
(M_P / M)^{(2p - 6)/7} \sim (M_P / M)^{2/7} $.  For $M \sim 10 $ TeV,
this is larger than $10^4$.  Note also that there is no plausible weak
coupling enhancement in this case, since 11D SUGRA has no dimensionless
expansion parameter.    The radial dilaton mass comes out of order
$M (M/M_P)^{(6/7)}$ and it cannot be seen in Cavendish experiments.

We have seen that there are a variety of radial stabilization schemes
whose plausibility depends on whether the underlying dynamics has stable
vacua with large flux, and a single example where we can get a
relatively low fundamental scale and a light radial dilaton with
fluxes of order one.
A much more exciting possibility in the case of bulk SUSY,
is simply that the radius is not
stabilized at all.  If the leading term in the potential is positive and
vanishes at large $R$ then the system is driven to infinity.
Thus, we might hope to explain a large current value for
$R$, simply as a result of the slow migration of $R$ to infinity
during the expansion of the three dimensional part of the
universe.  $R$ would be a form of quintessence
\cite{quintessence}.  Such a scenario would be a realization of Dirac's
old idea that Newton's constant is small because the universe is old.
Although we think that this is an attractive possibility, we will find
several problems with it \footnote{For an
alternative model of quintessence in
large radii models see ref.~\cite{benakli}.}.

We will assume that SUSY is broken on our
brane, at the scale $M$.  There is then a
cosmological constant of order $M^4$ which must
be canceled by fine tuning.  We must also require
that the additional, time dependent (because $R$ itself is time
dependent), potential
energy of $R$ is smaller than the conventional
contributions to the energy density of the universe
throughout most of cosmic history.  Today it
can be a finite fraction of closure density.

In order to achieve this goal, it is obviously
best to have a leading term in the potential which
is as high a power of $R^{-1}$ as possible.
 The $1/R^4$
term comes from the Casimir energy, and its
coefficient is expected to be of order one.  The
precise value of this coefficient depends on the
nature of the coupling of the massless bulk fields
to the SUSY breaking system on the wall.  For
the quintessence scenario, we must demand that
the coefficient be positive.

The bound that the current Casimir energy density
not exceed the critical density is
$R_0 > 10^{30} M_P^{-1} \sim 10^{-3} cm$.  Thus, in
the context of a theory with $M \sim 1$ TeV, and
$M_P \sim (R_0 M)^{n/2} M$ only
the scenario with two large dimensions is viable.
However, it is also hard to see how to avoid a term in the potential
coming from quadratic curvature terms, which scales like $(RM)^{-2}$.
In the presence of such a term, even the two dimensional case would seem
to be ruled out.  Let us agree to ignore this and explore
the other features of the radial quintessence scenario.

To determine whether the bound on the radial potential was
satisfied in the early history of the universe,
we invoke another bound, the constraint on the
time variation of Newton's constant.  This is
usually stated as ${\dot{G_N} \over G_N} < 10^{-11}/yr$.
This translates directly into a bound on the
time variation of $R$, which implies that $\dot{R} /R < 10^{-11}/yr.$.
Thus, during the time since nucleosynthesis
$RM$ has changed by at most a factor of order one.
On the one hand, this implies that there is
no problem with the energy density in $R$
dominating the universe at previous eras.  On
the other hand it means that our hope to
explain the large value of $R$ as a consequence
of the large age of the universe, cannot
work.  The initial value of $R$ cannot be so very
different than its value today.  This is certainly
disappointing.  A cosmological explanation for the large value of $RM$
would have to come from an inflationary era preceding the time of
nucleosynthesis.\footnote{We note a recent paper, \cite{dienesetal}, in which
an attempt is made to explain the large value of $RM$ in terms of
a thermal effective potential in a pre BBN era.}

It is worth pointing out that, in contrast to many models of
quintessence, the current model is not seriously constrained by
the cosmological variation of the fine structure constant. In this
model gauge fields live on the brane and do not couple directly to
$R$.  Their coupling comes only through radiative corrections
in which KK excitations are exchanged between the lines in a vacuum
polarization diagram.
The fine structure constant has the form \beq 1/\alpha
= 1/\alpha_0 + {\cal O}([RM]^{-q}) \eeq Thus \beq {\dot{\alpha}\over
\alpha} \sim {\dot{R}\over R} [RM]^{-q} \eeq Since we have already
established that $RM > 10^{14}$ and ${\dot{R}\over R} < 10^{-11} /yr. $, from
the bounds on the cosmological constant and the time variation of
$G_N$, this is less than the strongest bound on the time variation
of $\alpha$ from the Oklo natural nuclear reactor\cite{oklo}.

Another appealing feature of this picture is that the
scale of the potential is not necessarily connected
to the scale of supersymmetry breaking.  This is in
contrast to more conventional pictures\cite{dscosmo},
where these relations are tightly (and unacceptably)
constrained.

There probably is a fine tuning problem of order $10^{-2,3}$
coming from the effect of the $R$ field on astronomy.  In this
respect it behaves like a classic Brans-Dicke field.

None of these estimates touches on the question
of whether the dynamics of the $R$ field is
really compatible with the bounds.
That is, is the radial dilaton an acceptable dynamical model of
quintessence. We will not
attempt to answer this or any other cosmological question
in the current paper.

However, it is clear that in the best of all
possible scenarios, we can construct a viable
model of radial dilaton quintessence only
for $n=2$ and that even in that case the
large current value of the radius has to be
put in as an initial condition sometime before
the era of nucleosynthesis.  It remains to be seen
whether early universe cosmology and inflation
can give us a natural explanation of this number.

To summarize: models with SUSY in the bulk
can have similar stabilization mechanisms to non supersymmetric models.
Often the radial dilaton is very light in the SUSY case, and the models
are ruled out.  There is however a SUSY model with five large dimensions and
a $10$ TeV fundamental scale, which requires no large or small
parameters and has a radial dilaton which might be found in
sub-millimeter gravitational experiments.  The SUSY case also allows us
to construct a model in which the radial dilaton is not stabilized
and might act as a form of quintessence.  Compared to most quintessence
models, this scenario has less of a problem with the time variation of
the cosmological constant.
 However, the appealing idea of explaining the
large value of $R$ via the cosmological time variation of this
parameter runs into the observational bounds on the variation of
Newton's constant.  A viable model of $R$ as quintessence might be
constructed in the case $n=2$, but the explanation of its current value
would have to come from some very early inflationary era.

\section{Flavor, CP and Precision Electroweak Constraints}

A traditional  argument for a large fundamental scale has been the
absence of certain flavor changing processes. The most dramatic of
these is baryon number conservation. Unless the fundamental scale
is higher than around $10^{15}$ GeV,  proton stability must be
protected by symmetries.  However in many models of physics beyond
the standard model such as the MSSM, it is already necessary to
impose additional symmetries in order to suppress dangerous
renormalizable  and dimension 5  baryon  and lepton violating
operators. More elaborate symmetries can suppress baryon-number violating operators up to terms of
very high dimension.

Of course, there are many other sorts of flavor violation which
must be suppressed, and refs.~\cite{savasflavor,dvaliflavor} discussed
some aspects of this problem, and made an interesting
proposal. They suggested that the underlying theory might possess
some large flavor symmetry, and, in addition, several additional
branes, far from ``ours" on the scale $M$, but close when compared
to the scale of compactification. The hierarchy of quark and
lepton masses then arises because of a hierarchical separation of
the branes.  Explaining this hierarchy will raise many of the
issues discussed above, but it certainly provides a way of
parameterizing the breaking of chiral symmetries. Just as for the
bulk moduli we have discussed above, fixing these ``separation''
moduli at extreme values is problematic.  We have already argued
that there cannot be any approximate supersymmetry in these
pictures.  In weakly coupled string theory, there is no potential
between branes at the classical level.  However, there are states
whose mass grows with the separation of the branes. In
supersymmetric theories, there is a cancellation between bosonic
and fermionic modes, and perturbatively (and non-perturbatively,
if there is enough supersymmetry) there is no potential. However,
for non-supersymmetric theories, generically there is already a
force between branes already at the one loop level. In weakly
coupled string theory, at large distances,
this force corresponds to the exchange of
massless particles (gravitons, etc.) between the branes. Again,
then, one expects stabilization only for brane separations of
order the fundamental scale, if at all.

In the rest of this section, however, we will adopt the viewpoint
of refs.~\cite{savasflavor,dvaliflavor}, and suppose that the solution
of the flavor problems lies in a large separation of at least some
of the branes.  We will see that this still requires that the
fundamental scale be of order 6 TeV, and suppression of 
CP violating effects requires either  additional assumptions,   
a higher scale, or both.

In analyzing the effectiveness of this scheme in suppressing
flavor changing processes, we will impose a strict notion of
naturalness.  In particular, one expects in string theory that
operators permitted by symmetries are generated already at tree
level. In addition, in accord with the arguments of ref.
\cite{dineseiberg}, one
 doesn't expect
that the couplings should be weak.  As a result, operators allowed
by symmetries should be present at ${\cal O}(1)$ (one might argue
that they should be larger). With this assumption, one should
first examine constraints from {\it flavor conserving} operators,
which, due to   recent progress in precision electroweak physics,
can be quite severe. These come from processes such
as
\begin{itemize}
\item  Direct searches for 4-fermi couplings at LEPII.
\item  Atomic parity violation.
\item  Limits from precision  measurements of properties of the weak
gauge bosons on dimension 6 operators.

\end{itemize}
For instance,  data from the recently completed high luminosity
LEPII run place a 95\% C.L. limit on 4-lepton couplings such as
\beq {g^2\over M^2}\bar\ell_i \gamma^\mu\ell_i\bar\ell_j
\gamma_\mu\ell_j\ , \eeq
 of $M/g > 3 $ TeV \cite{lepc}.

A comparable or slightly stronger limit on $M/g$ can be found by
considering the effects of operators such as \beq {g^2 \over
M^2}\bar\ell_i \gamma^\mu\gamma_5\ell_i\bar q_j \gamma_\mu q_j\ .
\eeq on atomic parity violation \cite{apv}.

Other dimension 6 operators such as \beq {g^2 \over
M^2}(H^\dagger D^\mu H)(H^\dagger D_\mu H) \eeq can affect the
$\rho$ parameter. Requiring $\delta \rho/\rho < .003$~ places a
constraint $M/g >$~6~TeV. Similarly strong constraints on $M/g$
can be found by considering other precisely measured electroweak
observables such as the total width, total leptonic width, and
total hadronic width of the $Z$, which can be affected by
operators such as \beq {g^2 \over   M^2}(D^\nu Z_{\mu\nu})\bar
f_i\gamma^\mu f_i \eeq and \beq {g^2 \over   M^2}(H^\dagger D^\mu
H)\bar f_i\gamma_\mu f_i \ , \eeq where $f$ is any fermion.

In addition, any possible flavor symmetry involving the top quark
and left handed  bottom quark must be  maximally broken, and one
can also find constraints on $M/g $ from processes such as
\begin{itemize}
\item{}  The partial width $Z \rightarrow b + \bar b$, which is
affected by operators such as $\bar q_{3L}\gamma_\mu
q_{3L}H^\dagger D^\mu H$.
\item{}  $B_d \rightarrow \bar B_d$ mixing, which, after the effects
of CKM missing are considered, is affected by $\bar
q_{3L}\gamma_\mu q_{3L}\bar q_{3L}\gamma^\mu q_{3L}$.
\end{itemize}

These processes give a constraint on $M/g $ of about
 2 TeV.

We turn now to flavor violating processes in the first two
generations. Clearly, without some assumption of underlying flavor
symmetries, the scale $M$ is constrained to be far larger than a
few TeV. The authors of \cite{savasflavor} made a set of
assumptions which provide maximal suppression of unwanted
processes.  In the standard model, ignoring Yukawa couplings,
there is a global $U(3)^5$ symmetry. The first assumption is that
a large discrete subgroup of this symmetry is in fact a good
symmetry of the underlying theory.  (These chiral symmetries must
be discrete, in order to avoid light Nambu-Goldstone bosons on the
branes. Also, in M~Theory we do not expect global symmetries.)
 This symmetry is assumed
to be broken on branes which are far from our own (in units of
$M^{-1}$). There are some bulk fields which transform under these
symmetries and which communicate this breaking to our wall.  We
will denote these generically by $\chi$.  The simplest way to
suppress flavor changing neutral currents is to assume   that the
$\chi$ are in the representations necessary to give quark and
lepton masses, {\it i.e.} there are various $\chi^{u,d}$
transforming as $(3,3,1)$ and $(3,1,3)$, respectively, under the
discrete subgroups of $U(3)^q\times U(3)^u\times U(3)^d$, and
$\chi^{\ell}$ transforming as $(3,3)$ under $U(3)^\ell\times
U(3)^e$.  In this picture, some masses are smaller than others
because some of the branes are farther away than others, or some
of the $\chi$ are heavier than others (since the $\chi$
expectation values on our brane are exponentially suppressed by
the product of $\chi$ mass and the distance to the source of
$\chi$ vev).  For example, one can imagine a nearby wall
responsible for the mass of the $b$ quark, another, farther away,
for the $s$ quark, another for the electron mass, etc.  The
smallness of mixing angles could arise from the particular
alignment of the symmetry breaking on different walls, for example
(given the assumption of large discrete groups).

In such a picture, flavor violation is clearly suppressed by the
discrete subgroup of $U(3)^5$.  The question is by how much.  To
analyze  the amount of suppression of reasonably low dimension
operators, we assume the discrete flavor symmetry is large enough
so that we may proceed as if we have the full $U(3)^5$ symmetry.
First, it should be noted that CP conserving $\Delta s=2$
processes are reasonably safe, provided the lightest $\chi$ fields are
just those necessary to generate quark and lepton masses.   Dangerous
dimension six operators
can arise from terms such as \beq \label{qdop} {g^2 \over M^2}\bar
q_L \chi^d d_R \bar q_L \chi^d  d_R \ .
\eeq
Here $q$ refers generically to  the left handed quarks, $d$ to the right handed down-type quarks,   $\chi^d$  to   either the flavon field or to derivatives of the field with respect
to coordinates transverse to the brane. There is no reason to
expect any additional suppression of such derivatives, since the
mass of $\chi$ is presumably of order $M$. The quark masses also actually receive contributions from an infinite number of
operators involving $\chi$ and its derivatives. Thus   the
operator (\ref{deltastwo}) need not   be diagonal in the
down quark mass basis since in general neither derivatives of $\chi^d$ nor
the expectation value of $(\chi^d)^2$ can  be diagonalized
simultaneously with the expectation value of $\chi$. It is  however
true that in models such  as those suggested by
refs.~\cite{savasflavor,dvaliflavor}, the various entries of the
matrices indicated by $\chi^{u,d,\ell}$
will have the same order of magnitude
 as those of the corresponding Yukawa matrices. With these assumptions, in the
down quark mass basis, one could find a  $\Delta s=2$ operator of order
\beq
\label{deltastwo} {g^2 \over   M^2}\left({m_s^2
V_{cd}^2\over v^2}\right)\bar d_L    s_R \bar d_L   s_R . \eeq
The
matrix  element of the operator (\ref{deltastwo}) is enhanced by a
factor $m^2_K/m^2_s$ and by short distance QCD renormalization
group effects. However the real part of $K \bar K$ mixing is
adequately suppressed for $M/g$  of  900~GeV. Similar operators,
with $d$ replaced by $u$, can contribute to $D \bar D$ mixing.
However a $M/g$  of  order 1 TeV provides
 suppression consistent with current limits.

Other  $\Delta s=2$ operators arise  from terms such as \beq {g^2
\over   M^2} \bar q_L \chi \chi^\dagger   q_L \bar q_L \chi
\chi^\dagger q_L\eeq with various contractions of the indices.
But these are suppressed
  by four powers of $m_c$ or small CKM angles, and are less dangerous.
$\Delta s=1$ operators also provide weaker limits.

Consideration of CP violating operators provides, potentially,
more stringent constraints.  To explain the smallness of CP
violation, CP must be a good symmetry of the bulk, violated
spontaneously  on a distant brane. Otherwise, unless the scale $M$
is very big,  CP violating operators     such as
\beq \label{ggg}
{g^2 \over   M^2} f_{abc} 
 G_a{}^\mu{}_\nu  
 G_b{}^\nu{}_\lambda 
   \tilde G_c {}^\lambda_{\mu} \eeq
will make huge contributions to the dipole
moment of the neutron $d_n$.

Assuming the CKM mechanism of CP violation, it is necessary that
CP be violated on some of the branes responsible for quark masses.
An order one CKM phase is not possible if CP is only violated on
the branes responsible for the first generation masses. If CP is
violated generically on the   the branes which provide the   quark
masses,   then operators such as (\ref{deltastwo}) and also such as
\beq \label{edm} {g^2\over M^2} e F_{\mu\nu} \bar q_L
\chi^d\sigma^{\mu\nu}d_R H\eeq (where again $\chi$ may refer to a
derivative of itself, not necessarily real in the same basis as
the quark masses) may have complex coefficients. The former
potentially gives too large an $\epsilon_K$ unless $M/g>$~10~TeV.
The latter may give  a too large $d_n$ unless $M/g>
40$~TeV (assuming the contribution of the down quark dipole
moment to $d_n$ is given by naive dimensional analysis
\cite{nda}). The contributions of the latter must be substantially suppressed if the scale $M$ is anywhere near the electroweak scale.  One way to suppress the contribution of the operator
(\ref{edm}) is if the $\chi$ field is rather light compared with
$M$ so that contributions of its derivatives are suppressed.  It is also quite possible that even if $\chi^d$ is not light, its derivatives are real and diagonal in the same basis as $\chi^d$. Recall that we are assuming that the bulk physics preserves a
large {\it discrete} subgroup of the flavor symmetries. Thus the bulk physics provides a potential for the matrices $U(L)$ and $U(R)$ which diagonalize $\chi^d$, which is minimized for discrete values. It is thus plausible that these matrices are not spatially varying,
since to do so might cost too much potential energy. There would
therefore be a basis, in which the down masses were real and diagonal,
and CP and strangeness were good symmetries until the effects of
$\chi^u$ (and its derivatives) are considered.
However it will still be expected that, {\it e.g.} $\langle
\chi^u
{\chi^u}^\dagger\chi^d\rangle$ will be complex in the quark mass basis and so
operators such as \beq \label{edmII} \left({g^2\over M^2}\right) e
F_{\mu\nu} \bar q_L \chi^u
{\chi^u}^\dagger\chi^d\sigma^{\mu\nu}d_R H\eeq give a  contribution to $d_n$. The contribution from the down quark is not dangerous, and even if the strange quark matrix elements are order 1 as given by naive dimensional analysis and indicated by several 
experiments~\cite{sme}, then one would
have to have $M/g>$~1~TeV in order to suppress
the contribution to $d_n$ from the strange quark electric dipole moment.

Note that suppression of the contribution of the operator (\ref{edm}) still does not
completely explain the small size of $d_n$ which still could arise
{}from the  strong CP parameter $\bar\theta$ \footnote{This strong
CP problem might still be solved by an invisible axion in the
bulk, see ref. \cite{savasconstraints}, or a massless
up quark.}. More restricted assumptions about CP violation can
ameliorate this problem as well as the constraints previously
mentioned. 
One alternative is that there is no CP
violation either in the bulk or on the branes which serve as sources for
$\chi^{u,d}$. Then the quark mass matrix is nearly real and the
phase in the CKM matrix   is very small\footnote{Unlike in the
  standard model case\cite{realCKM}, current data still allows a real
  CKM matrix if there are nonstandard contributions to $B \bar B$
  mixing.}. 
This could have the
advantage of solving the strong CP problem. However, one  then
must hypothesize more complicated mechanisms
to provide CP violation in $K-\bar K$ mixing. For instance there
could be another distant wall, on which CP and, say,  the flavor
symmetry  acting on the left handed quarks is broken, but the
other flavor symmetries are conserved. Thus this wall cannot serve as a source for $\chi^{u,d}$. The expectation value of a
heavy $\chi^q$ field transforming as a  27 under the $SU(3)$ of the
left handed quarks could provide a $\Delta s=2$ CP violating
operator
\beq
\bar s_L\gamma_\mu d_L \bar s_L\gamma^\mu d_L \ .
\eeq
Quark electric dipole moments would now arise only from additionally suppressed operators such as
\beq
\label{edmIII}
{g^2\over M^2} e F_{\mu\nu}
\bar q_L \chi^q \chi^u{\chi^u}^\dagger\chi^d\sigma^{\mu\nu}d_R H\ .
\eeq
The strong CP
parameter $\bar\theta$ and dangerous CP violating operators such
as (\ref{ggg}) could be sufficiently small provided that any
$\chi$ fields which  combine into a complex flavor singlet have enough suppression of the product of their expectation values.   Within a few years such a solution will
be definitely tested by the B factories, which might provide
direct evidence for a nonzero CKM phase.

 Lepton number  and lepton flavor violation also must be
highly suppressed. It is not reasonable simply to assume that the
individual lepton flavors are conserved  since there is good
evidence for violation of lepton flavor in neutrino oscillations.
One might assume that  small  Dirac neutrino masses arise from the
mechanism of  ref. \cite{savasflavor,numass}.
Neutrino masses then imply   large violation of the $U(3)$
symmetries of the left handed leptons, which we parameterize by $\chi^\nu$. Lepton flavor violation
{}from higher dimension operators such as
\beq
{g^2\over M^2}\bar\ell_L \chi^\nu{\chi^\nu}^\dagger\gamma_\mu\ell_L
\bar\ell_L \gamma^\mu\ell_L\ ,
\eeq
 could lead to visible nonstandard decays such as $\tau\rightarrow 3\mu$ or $\mu\rightarrow 3e$ unless $M/g$ is very large. However by choosing different  properties for the right
handed neutrinos or a different mechanism for neutrino masses one
clearly  has the option of assuming that lepton flavor
violating $\chi$ expectation values are very small and not
dangerous\footnote{For another discussion of neutrino masses and large
  extra dimensions see ref.~\cite{Faraggi}.}. 
Because there
are  many possible options for the neutrino masses, we do not
consider lepton flavor violating constraints further in this
paper.

However even with suppression of lepton flavor violation, a
constraint comes from a possible contribution to the
anomalous magnetic moment of the muon\footnote{A similar bound may also be
  placed by considering the explicit contributions from KK 
modes \cite{graesser}.}, from
\beq 
{g^2 \over   M^2 } 
e F_{\mu\nu}\bar\ell_L \chi^\ell \sigma_{\mu\nu}e_R H\ .
\eeq
This is too large unless $M/g> 1$~TeV.

We have seen that   consideration of the effects of flavor
conserving higher dimension  operators   suggest that
the fundamental scale should be at least 6 TeV, and in many models of
CP violation the scale must be at least 10 TeV. Also a peculiar
flavor symmetry 
is required which   must be very judiciously   broken. These scales are
somewhat troubling from the perspective of understanding the
lightness of the Higgs particle, which requires some mechanism to
suppress its mass squared term to   of order
$(200~{\rm GeV})^2$---$(800~{\rm GeV})^2$.
If this small term arises by accident,   a fine tuning  is
required which is greater than at least a part in  100 .  Of course, it
is possible that for some mysterious reason  the natural size of
the Higgs mass is not $M$.  There might  be   some approximation
in which the Higgs is light, and receives its mass radiatively.
  Also, there might be very
small couplings that enter in the higher dimension operators.
However avoiding substantial fine tuning of the weak scale clearly
places additional  nontrivial constraints on the underlying theory.

\section{Cosmology}

The cosmology of theories with several large dimensions could
potentially be quite rich.  At very early times, the gravitational
and gauge couplings could be far from their present values.  The
conventional horizon and flatness problems might take a quite
different form, and might be amenable to quite different
solutions, than usually assumed.

Even so, cosmology is likely to pose serious problems for theories
with such light moduli.  Lacking a detailed cosmological model, we
will content ourselves with a few brief remarks in this section.

First, as noted in \cite{savasconstraints}, it is necessary
that the bulk moduli are in their ground states to a high
degree of accuracy at early times.  Indeed, these authors define a
``normalcy temperature", $T_n$, above which thermal production on
the branes will over-produce the bulk modes.  This temperature is
quite low. In the case of two dimensions
it is barely above nucleosynthesis temperatures.
So it is necessary that ``inflation" only reheat to temperatures
very slightly above an MeV.  In all cases, the temperature is
orders of magnitude below the fundamental scale.

We find this
condition quite puzzling.  For example, if the inflaton is a bulk
modulus, it will contribute to the bulk cosmological constant and
inevitably displace the radial dilaton far from its true minimum.
On the other hand, without inflation in the bulk, it is hard to
understand how the bulk system got into its ground state.  So one must
have inflation in the bulk with a reheat temperature below an MeV.
Furthermore, in order to account for the excitation of the standard
model during nucleosynthesis, the bulk inflation (or some other
mechanism) must, as discussed in \cite{savasconstraints},
leave over some excited field on the brane which dumps its energy
almost entirely into standard model degrees of freedom.
This seems particularly difficult in scenarios with only two large
extra dimensions.  In this case, homogeneous excitations on the brane
have logarithmically growing couplings to bulk modes\cite{ab}. Finally,
assuming that all of these other criteria have been met,
the reheat temperature on the brane must be lower than the normalcy
temperature of \cite{savasconstraints}.

The potential difficulty
posed by this last constraint is illustrated by the
otherwise attractive model of inflation proposed in
\cite{dvalitye} (another model for brane inflation appears in
\cite{ovrut}).  These authors make the very interesting point
that in the brane scenario there are natural candidate inflatons.
These are the fields which describe the separation of the branes.
These fields have potentials, arising from massless exchanges,
which fall rapidly to zero when the branes are separated.  When
the branes are nearby, they are expected to have potentials with
curvature of order $M$.  Thus if the ground state has some branes
close together, and yet they start out well separated at early
times, the system can inflate.  However, there are at least two
difficulties with such a picture.  First, as pointed out by the
authors, it is difficult to have sufficiently large fluctuations.
Second, the natural reheating
temperature is of order $M$.  Smaller scales seem to require fine
tuning.

Finally, particularly for the case $n=2$, there are potentially
efficient production mechanisms for the bulk modes , which have not been
carefully studied.  Suppose for example that the universe undergoes a
phase transition on a time scale small compared to $R_0$.  This would be
the case in the brane separation transition described above.  Recall
that for $n=2$ there is enormous energy stored in the gravitational
field surrounding the brane, spread over a millimeter.  If the
transition is too rapid, this energy cannot be dissipated adiabatically;
it will be principally radiated in bulk modes.

 We are not sure that these problems are insurmountable, and existing
 models of inflation also have their
 difficulties.  Still, absent a concrete model,
 one is entitled to be skeptical of the
possibility of arranging such a delicate sequence of events
in the early universe.   It is interesting that models with dimensions
of order a (10 MeV)$^{-1}$ or smaller
avoid most of the difficulties with Big Bang
Nucleosynthesis, because the KK modes are unexcited at nucleosynthesis
temperatures.  This is another reason why we consider this the most
plausible realization of the large extra dimension scenario.

Another issue that will have to be resolved in these theories is an
analog of the cosmological moduli problem.  We have argued quite
generally that the radial dilaton will be a field with gravitational
couplings, mass of order $10^{- (3\ or\ 4)}$ eV, and a potential energy
density of order a (TeV)$^4$ .  Thus, a mechanism
for setting this field at its minimum before or during inflation must be
found in order to avoid a matter dominated universe at nucleosynthesis
energies.   In scenarios with two large dimensions there will be similar
problems with the KK modes.

These problems appear formidable to us, but the cosmology of brane
worlds has many potential sources of surprise.  The interplay of
inflation on and off the brane and a rich spectrum of energy scales
seems quite complicated.

\section{Conclusions}

The possibility that the fundamental scales of nature are
comparable to the electroweak scale, while the scales of
compactification are large, is extremely exciting.  In this note
we have argued that while this possibility is not ruled out by any
phenomenological considerations, it is highly constrained, particularly
within the framework of M~Theory.    We argued that purely
phenomenological constraints, coming from precision electroweak
measurements as well as flavor violating rare processes, push the
lowest allowed value of the fundamental scale up to $6 - 10$ TeV.
We believe that we have been very conservative in deriving these bounds,
and used ameliorating assumptions proposed by other authors
with a large degree of faith.  In particular, although the authors
of \cite{savasflavor} propose a resolution of the flavor problem
employing large discrete nonabelian symmetries, broken on distant
branes, no explicit models with all the required properties have
been constructed\footnote{For a slightly different and  more
  explicit flavor proposal see ref.~\cite{tevsm}.}.  Our analysis assumed that models incorporating these
ideas will eventually appear.   If not, the flavor constraints are
probably stronger.   We have made similar, maximally mitigating,
assumptions with respect to CP violation.

 Probably the most severe problems with models of large dimensions
were associated with the stabilization of the radius.  In the
non-supersymmetric case, it is necessary that there be large,
quantized fluxes of magnitude at least $10^5$.
It is not clear whether M~Theory allows such large
fluxes in spaces which are Minkowski space times a compact manifold.
When the bulk theory obeys SUSY,
we saw that if there are five large dimensions, and
one is willing to push the fundamental scale up to at least $10$ TeV
(which anyway one must do to satisfy constraints from precision
experiments),  while keeping the boundary cosmological constant
scale fixed at $1$ TeV ({\it e.g.} by invoking SUSY on the brane),
then one can stabilize the radius with moderate
values of the flux.  The combination of experimental constraints
and plausible stabilization mechanisms suggest that this kind of model
is the most likely realization of large dimension scenarios.
There are many other SUSY scenarios which require large values of flux
and have radial dilaton masses which contradict experiment.

An interesting possibility, which we had hoped would provide more motivation
for the idea of large dimensions, is that the radial dilaton might
play the role of the quintessence field.  In the case of two large
dimensions, the scales at least are plausible:  assuming a
non-zero Casimir potential, the mass of the radial dilaton is
naturally within a few orders of magnitude of the present value of
the Hubble constant.  However, one must fine tune the coefficient of a
term in the effective action quadratic in curvatures (or find a model
with vanishing bulk curvature) in order to have the Casimir energy
dominate the potential.
The cancellation of the cosmological
constant between boundary and bulk effects is still mysterious,
but at least the value of the radius dependent terms is of the
right order of magnitude. There are two problems with this idea. The
Brans-Dicke coupling of the radial dilaton is
naturally of order one, whereas observation restricts its value to
$10^{-4}$ or so.   And bounds on the time variation of Newton's constant
imply that one cannot explain the current large value of the radius
as a consequence of the evolution of the universe during the long period
of time since BBN.
Still, given that the rest of the numerology is
so suggestive, and also the high degree of fine tuning required by
existing quintessence models\cite{koldalyth}, it is probably worth
exploring this intriguing idea further.  To do so, we would have to
understand brane world models during inflationary eras.   The analysis
of such scenarios is only just beginning.

There are in fact a large number of interesting questions about large
dimension scenarios that can only be understood in the context of
inflationary cosmology.   In particular, for those values of $n$ for
which the bulk KK spectrum is below $1$ MeV, the initial conditions one
must assume in order to make the model consistent with BBN are quite
bizarre.   The brane is excited to a temperature above $1$ MeV while the
bulk is in its ground state.  This could only be accounted for by some
inflationary mechanism which put both bulk and boundary into their
ground states apart from some stretched scalar field on the
brane\footnote{T.B. thanks Nima Arkani-Hamed for explaining this
scenario to us.}.  This scalar must couple strongly to the standard
model in order to dump most of its energy on the brane.  On the other
hand, its reheat temperature must be low (not more than a few GeV)
in order to avoid excitation of
the bulk through its well understood couplings to the standard model.
Furthermore, baryogenesis and structure formation must all be squeezed
into this rather abbreviated cosmic history.   The absence of coupling
to the bulk is particularly hard to understand in models with two large
dimensions, where homogeneous excitations on the brane give rise to
logarithmically growing effects in the bulk.

The inflationary cosmology of brane worlds is only beginning to be
explored \cite{dvalitye} and it remains to be seen whether models
which meet all the challenges can be constructed.  Here we note
(once again) only that many of these problems are ameliorated in
scenarios with a large number of dimensions with inverse size of
order $10$ MeV.  The KK modes are now above nucleosynthesis
temperatures.   Thus, from the cosmological point of view as well,
models with a large number of large dimensions and a $10$ TeV
fundamental scale seem like the most plausible realization of
large dimension scenarios.   These models could have exciting
phenomenology both in gravitational and accelerator experiments
and we believe they deserve further study.

\noindent
{\bf Acknowledgements:}

\noindent
We thank Sean Carroll, S. Dimopoulos and Nima Arkanani-Hamed
for discussions.   This work of M.D. was supported in part by the U.S.
Department of Energy.  The work of T.B. was supported in part by the
Department of Energy under grant number DE-FG02-96ER40559. The work of A.N. was supported in part by the Department of Energy under 
grant no. DE-FG03-96ER40956. A.N. would like to thank the UCSC and CERN theory groups for hospitality.


\end{document}